\documentclass[final,3p,times]{elsarticle}

\usepackage{amssymb}
\usepackage{amsmath}

\usepackage{lineno}

\journal{Journal of Subatomic Particles and Cosmology}

\begin{document}

\begin{frontmatter}



\title{The role of strangeness and baryon enhancement in heavy-quark hadronisation from pp to Pb--Pb collisions with ALICE}

\author{Fabrizio Chinu\textsuperscript{a,b} for the ALICE Collaboration}
\affiliation[aaa]{organization={Università degli Studi di Torino},
             addressline={Via Pietro Giuria 1},
             city={Torino},
             postcode={10125},
             state={TO},
             country={Italy}}

 \affiliation[bbb]{organization={Istituto Nazionale di Fisica Nucleare, Sezione di Torino},
             addressline={Via Pietro Giuria 1},
             city={Torino},
             postcode={10125},
             state={TO},
             country={Italy}}

\begin{abstract}
In this contribution, a combined study of charm-meson and charm-baryon production with ALICE is presented. The $\Lambda_{\mathrm{c}}^{+}/\mathrm{D^{0}}$ ratio in pp collisions at $\sqrt{s} = 13.6$ TeV and the first measurement of the nuclear modification factor of $\Xi_{\mathrm{c}}^{0}$ baryons in Pb--Pb collisions at $\sqrt{s_\mathrm{NN}} = 5.02$ TeV are reported. Furthermore, measurements of the $\mathrm{D_s^+/D^+}$ production ratio from low-multiplicity pp collisions at $\sqrt{s} = 13.6$ TeV to Pb--Pb collisions at $\sqrt{s_\mathrm{NN}} = 5.36$ TeV are presented. Lastly, the first measurement of the $p_\mathrm{T}$-differential $\mathrm{D_{s1}(2536)^+/D_s^+}$ yield ratio is shown.

\end{abstract}




\end{frontmatter}



\section{Introduction}
\label{sec:intro}
Hadronisation is a fundamental process of nature, in which quarks combine to form hadrons. It involves non-perturbative QCD effects, thus making first-principles calculations challenging. Instead, this process is usually parametrised using data collected in $\mathrm{e^+e^-}$ and ep collisions, where hadronisation is assumed to occur via in-vacuum fragmentation, and used to describe hadron production in proton--proton (pp) collisions, under the assumption of a universal hadronisation mechanism. Measurements of charm baryon-to-meson ratios in pp collisions at the LHC~\cite{ALICE:2022exq}, however, show a significant enhancement compared with those in $\mathrm{e^+e^-}$ collisions, challenging the assumption of universal hadronisation and motivating models incorporating modified hadronisation mechanisms, such as coalescence, or the formation of small droplets of quark-gluon plasma. Measurements of the production of strange hadrons as a function of charged-particle multiplicity in pp collisions~\cite{ALICE:2016fzo} show an increasing trend with multiplicity, which hints at the possible formation of quark-gluon plasma also in small systems. In this context, measurements of strange and non-strange charm hadrons across collision systems are crucial to probe the hadronisation mechanism of charm quarks. In particular, hadronisation via coalescence in a strangeness-rich environment, such as the one created in Pb--Pb collisions, is expected to enhance the production of strange charm hadrons.
\section{Measurements of charm baryons in pp and Pb--Pb collisions}
\label{sec:baryons}

\begin{figure}[tb]
    \includegraphics[width=0.42\textwidth]{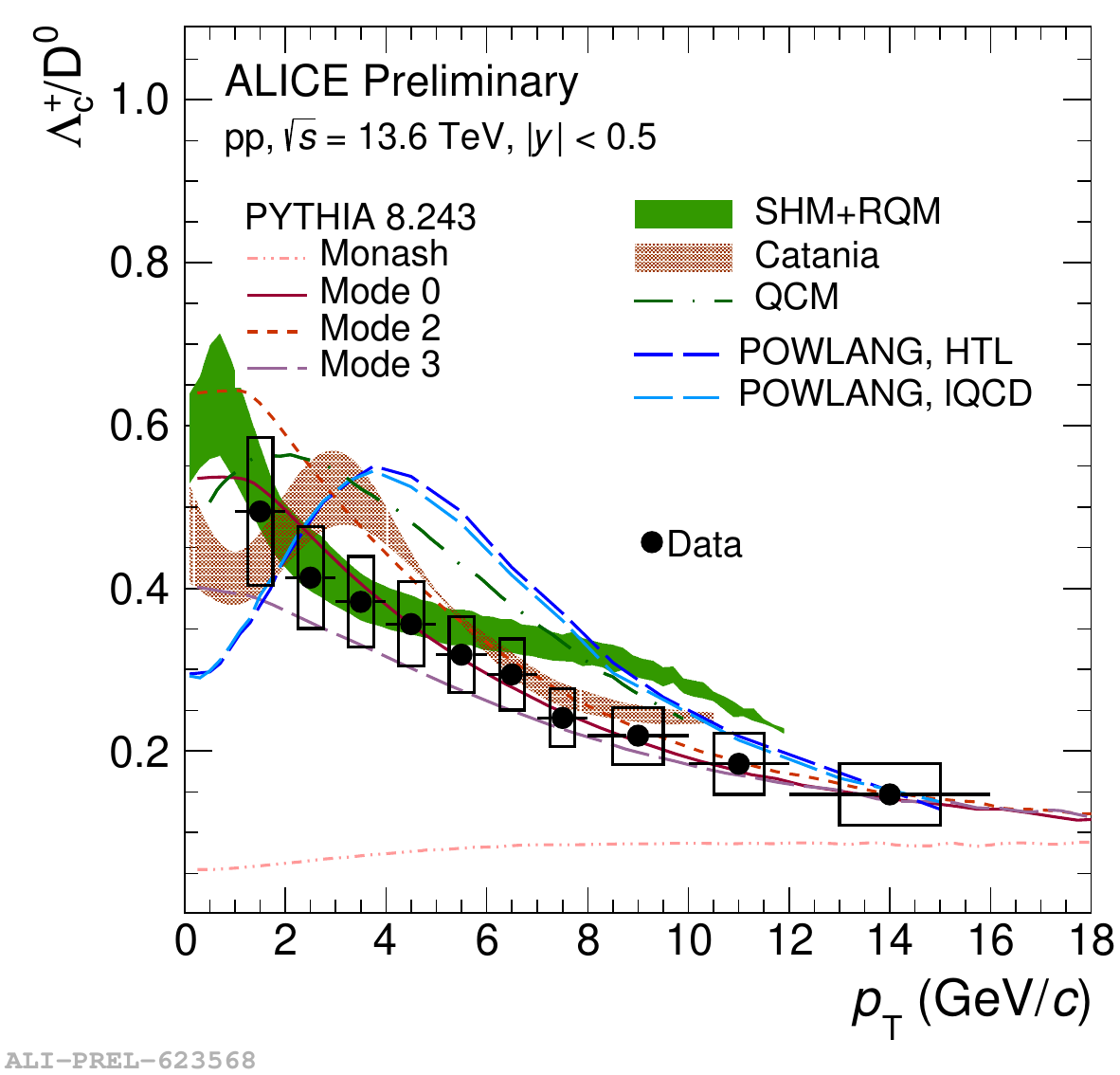}
    \includegraphics[width=0.56\textwidth]{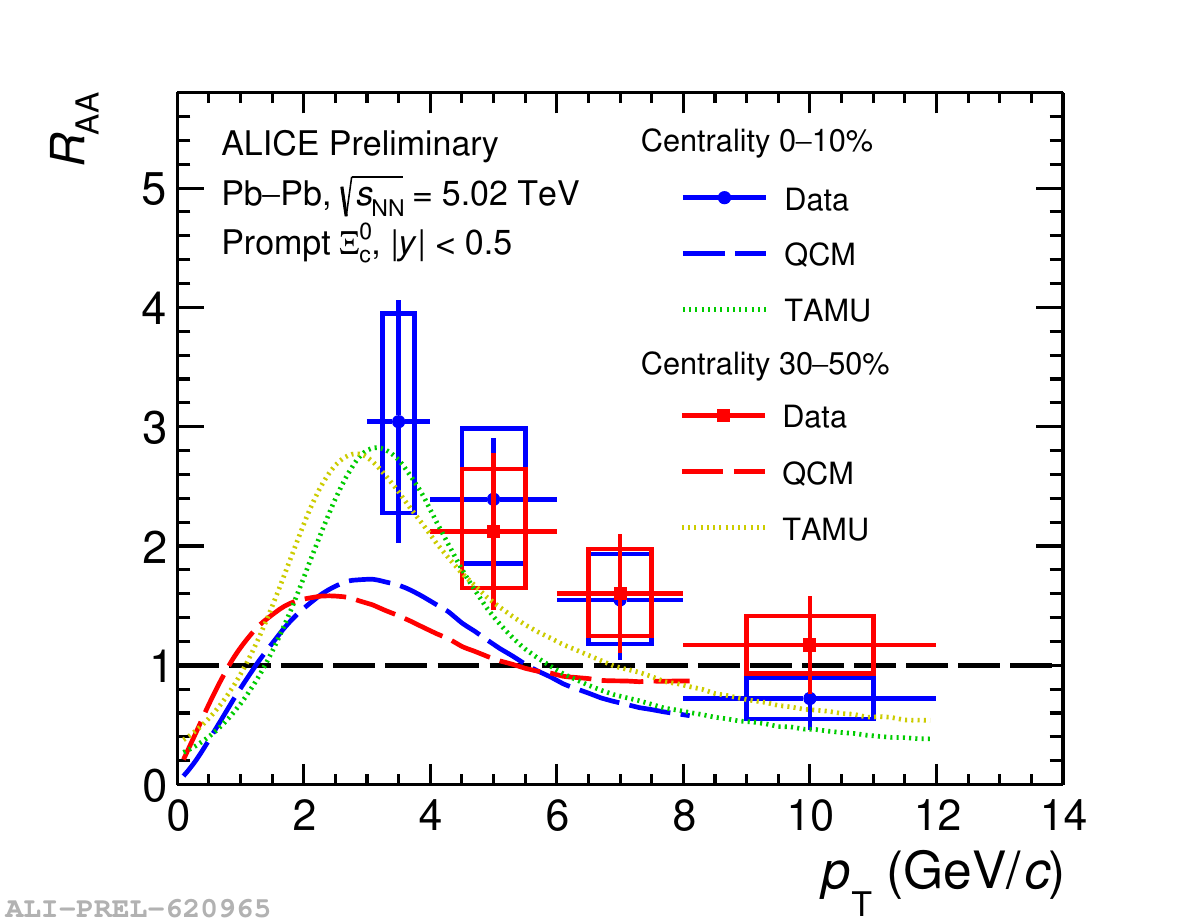}
    \caption{Left panel: $\Lambda_{\mathrm{c}}^{+}/\mathrm{D^{0}}$ ratio measured in pp collisions at $\sqrt{s} = 13.6$ TeV, compared with \textsc{Pythia 8} with the Monash tune~\cite{Skands:2014pea}, \textsc{Pythia 8} with colour reconnections beyond the leading-colour approximation~\cite{Christiansen:2015yqa}, Catania~\cite{Minissale:2020bif}, QCM~\cite{Song:2018tpv}, POWLANG~\cite{Beraudo:2023nlq}, and a statistical model with additional excited charm states predicted by the Relativistic Quark Model (SHM+RQM)~\cite{Ebert:2011kk, He:2019tik}. Right panel: $R_\mathrm{AA}$ of $\Xi_{\mathrm{c}}^{0}$ baryons measured in Pb--Pb collisions at $\sqrt{s_\mathrm{NN}} = 5.02$ TeV in the 0--10\% and 30--50\% centrality intervals, compared with QCM~\cite{Song:2018tpv} and TAMU~\cite{He:2019vgs} models.}
    \label{fig:baryons}
\end{figure}

Measurements of charm baryon-to-meson production-yield ratios, such as $\Lambda_{\mathrm{c}}^{+}/\mathrm{D^{0}}$, provide important constraints on charm hadronisation mechanisms. The left panel of Fig.~\ref{fig:baryons} shows the $\Lambda_{\mathrm{c}}^{+}/\mathrm{D^{0}}$ ratio measured in pp collisions at $\sqrt{s} = 13.6$ TeV. At low $p_\mathrm{T}$, the ratio is significantly larger than the LEP average of $0.113 \pm 0.013 \pm 0.006$~\cite{Gladilin:2014tba} and than the predictions of models in which fragmentation is tuned to measurements at $\mathrm{e^+e^-}$ colliders, such as \textsc{Pythia 8} with the Monash tune~\cite{Skands:2014pea}. A better description of the data is provided by models incorporating additional hadronisation mechanisms, such as colour reconnections beyond the leading-colour approximation~\cite{Christiansen:2015yqa} in \textsc{Pythia}, or hadronisation via coalescence in Catania~\cite{Minissale:2020bif}, QCM~\cite{Song:2018tpv}, and POWLANG~\cite{Beraudo:2023nlq}, or an additional set of yet-unobserved excited charm states predicted by the Relativistic Quark Model~\cite{Ebert:2011kk} in a statistical hadronisation model (SHM+RQM)~\cite{He:2019tik}. At high $p_\mathrm{T}$, the $\Lambda_{\mathrm{c}}^{+}/\mathrm{D^{0}}$ ratio approaches the value measured in $\mathrm{e^+e^-}$ collisions, suggesting that fragmentation becomes the dominant hadronisation mechanism in this regime.

Possible modifications of the hadronisation mechanism in Pb--Pb collisions are investigated by measuring the nuclear modification factor $R_\mathrm{AA}$ of charm hadrons. The right panel of Fig.~\ref{fig:baryons} shows the first measurement of $R_\mathrm{AA}$ of $\Xi_{\mathrm{c}}^{0}$ baryons, performed by the ALICE Collaboration at midrapidity ($|y|<0.5$) in Pb--Pb collisions at $\sqrt{s_\mathrm{NN}} = 5.02$ TeV. The $R_\mathrm{AA}$ reaches a value of $\approx 3$ at its maximum, the largest among the measured charm hadrons, suggesting a strong enhancement in the production of strange charm baryons in Pb--Pb collisions. The measurement is compared with predictions from the QCM~\cite{Song:2018tpv} and TAMU~\cite{He:2019vgs} models, which implement hadronisation via coalescence. Both models qualitatively describe the trend observed in the data, despite predicting a smaller enhancement than measured.

\section{Strangeness enhancement of charm mesons from pp to Pb--Pb collisions}
\label{sec:mesons}

\begin{figure}[tb]
    \center
    \includegraphics[width=0.8\textwidth]{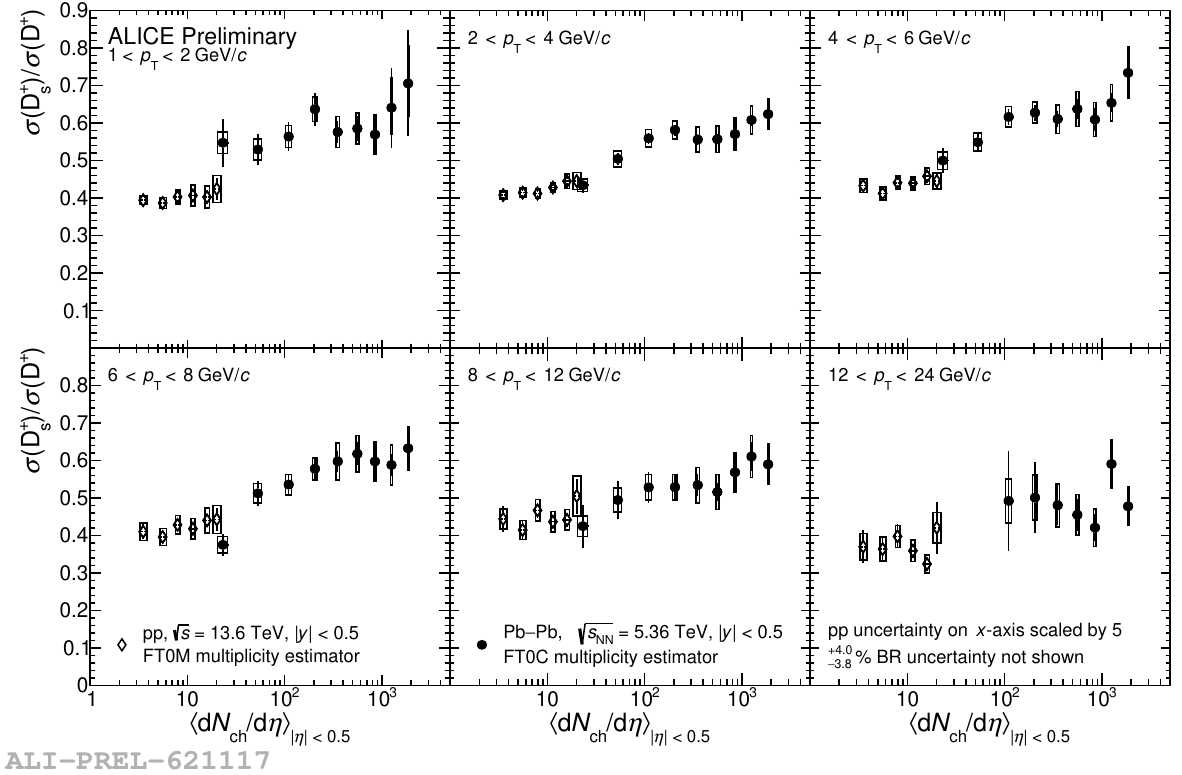}
    \caption{$\mathrm{D_s^+/D^+}$ ratio measured by the ALICE Collaboration at midrapidity ($|y|<0.5$) in pp collisions at $\sqrt{s} = 13.6$ TeV and in Pb--Pb collisions at $\sqrt{s_\mathrm{NN}} = 5.36$ TeV as a function of charged-particle multiplicity, for different intervals of $p_\mathrm{T}$.}
    \label{fig:ds_over_dplus_vs_pt}
\end{figure}

The ratio of strange-to-non-strange charm mesons, $\mathrm{D_s^+/D^+}$, is a sensitive probe of charm-quark hadronisation. Figure~\ref{fig:ds_over_dplus_vs_pt} shows the $\mathrm{D_s^+/D^+}$ ratio measured by the ALICE Collaboration at midrapidity ($|y|<0.5$) in pp collisions at $\sqrt{s} = 13.6$ TeV and in Pb--Pb collisions at $\sqrt{s_\mathrm{NN}} = 5.36$ TeV as a function of charged-particle multiplicity, for different $p_\mathrm{T}$ intervals. In pp collisions, the ratio shows a flat trend as a function of multiplicity, as expected from a scenario of hadronisation via fragmentation only. In contrast, in Pb--Pb collisions, the ratio increases with multiplicity for $p_\mathrm{T} < 8$ GeV/$c$, suggesting that coalescence plays a role in charm hadronisation in Pb--Pb collisions. At high $p_\mathrm{T}$ ($> 8$ GeV/$c$), the ratio in Pb--Pb collisions is compatible with the one measured in pp collisions, indicating that fragmentation becomes the dominant hadronisation mechanism at large $p_\mathrm{T}$.

\begin{figure}[tb]
    \includegraphics[width=0.476\textwidth]{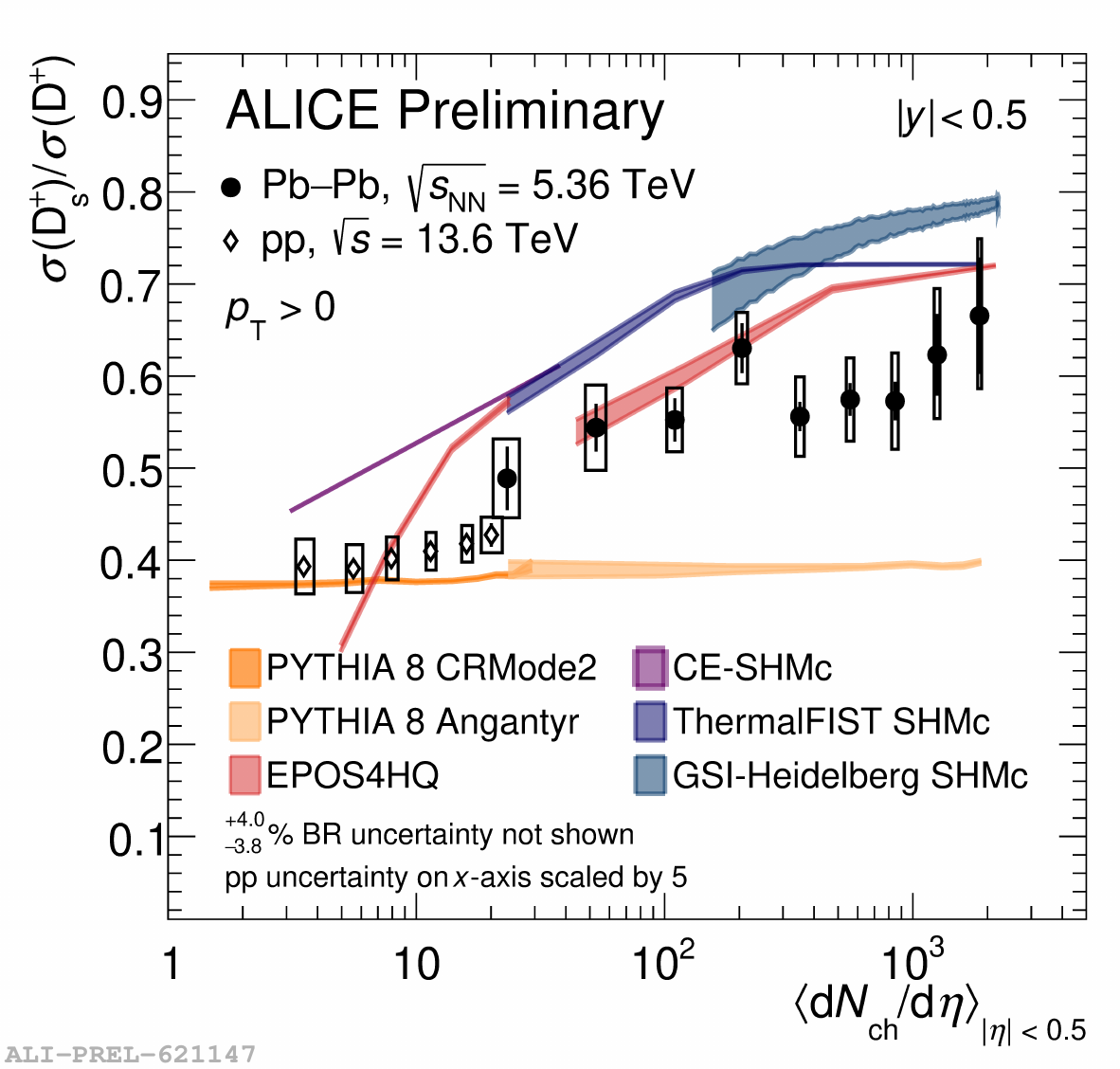}
    \includegraphics[width=0.455\textwidth]{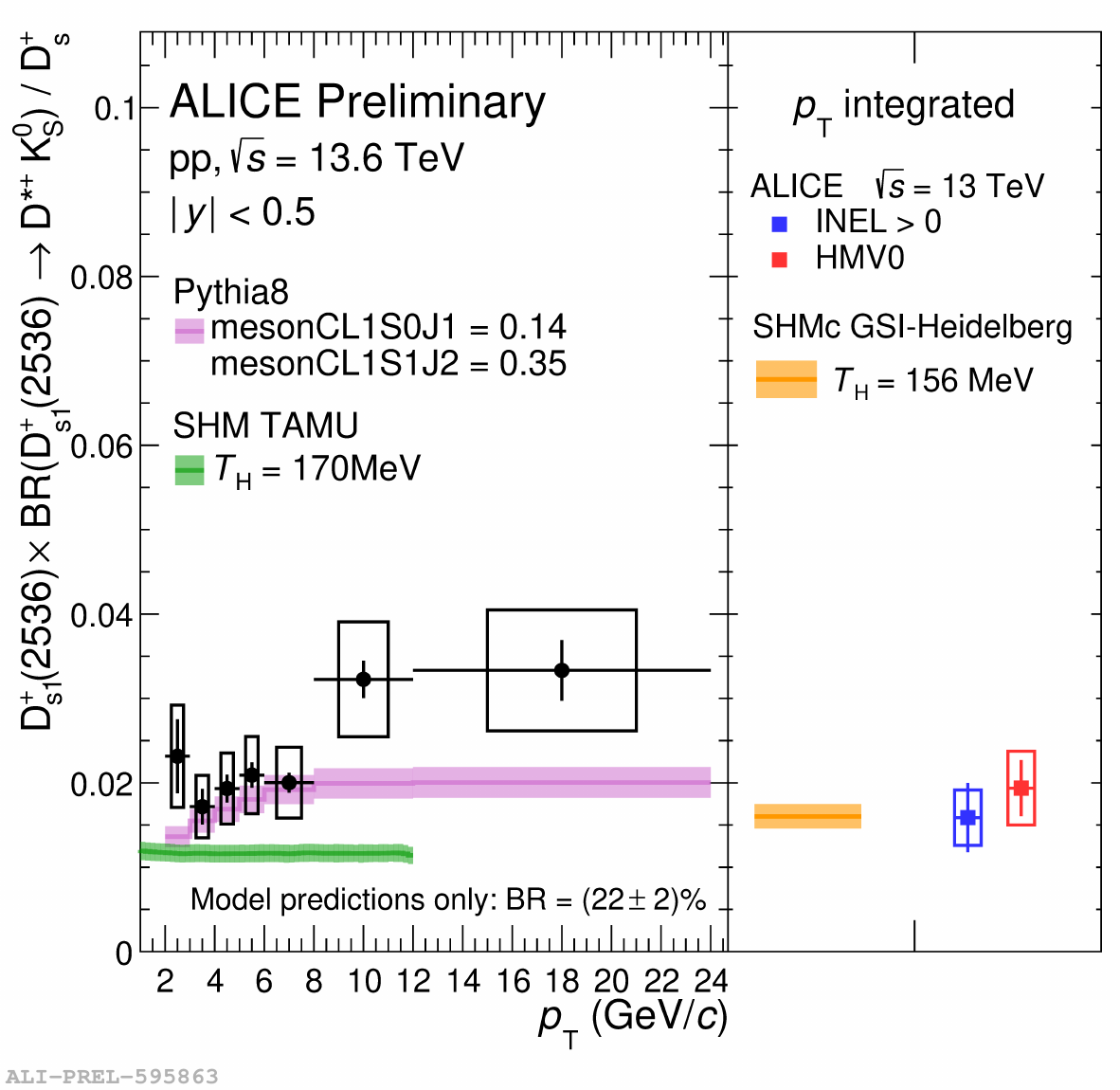}
    \caption{Left panel: $p_\mathrm{T}$-integrated $\mathrm{D_s^+/D^+}$ ratio as a function of charged-particle multiplicity in pp collisions at $\sqrt{s} = 13.6$ TeV and in Pb--Pb collisions at $\sqrt{s_\mathrm{NN}} = 5.36$ TeV, compared with \textsc{Pythia 8} with colour reconnections beyond the leading-colour approximation~\cite{Christiansen:2015yqa, Bierlich:2018xfw}, EPOS4HQ~\cite{Zhao:2024ecc}, and statistical hadronisation models~\cite{Andronic:2021erx, Vovchenko:2019pjl, He:2019tik}. Right panel: $p_\mathrm{T}$-differential $\mathrm{D_{s1}(2536)^+/D_s^+}$ production yield ratio multiplied by the \mbox{$\mathrm{D_{s1}(2536)^+} \rightarrow \mathrm{D^{*+}} \mathrm{K_S^0}$} branching ratio measured in pp collisions at $\sqrt{s} = 13.6$ TeV, compared with \textsc{Pythia 8} with colour reconnections beyond the leading-colour approximation~\cite{Christiansen:2015yqa} and statistical hadronisation models~\cite{Andronic:2021erx, He:2019tik}.}
    \label{fig:mesons}
\end{figure}

To assess whether an increase in the total production of strange-charm mesons as compared to non-strange ones is observed, or whether the observed trend with charged-particle multiplicity is due to a different transverse momentum distribution in Pb--Pb collisions, the $\mathrm{D_s^+/D^+}$ ratio was integrated over $p_\mathrm{T}$ and is shown in the left panel of Fig.~\ref{fig:mesons}. An increase in the measured ratio is observed in Pb--Pb collisions as compared to pp collisions, suggesting that hadronisation via coalescence may occur in this system. The measurement is compared with expectations from \textsc{Pythia 8} with colour reconnections beyond the leading-colour approximation~\cite{Christiansen:2015yqa, Bierlich:2018xfw}, which predicts a flat trend as a function of multiplicity, also in Pb--Pb collisions. EPOS4HQ~\cite{Zhao:2024ecc} is a core-corona model, where in the corona region hadronisation occurs via fragmentation, while in the core region coalescence can take place. The model predicts an increase of the ratio with multiplicity, qualitatively describing the trend observed in the data, but it also predicts a dependence on multiplicity in pp collisions, which is not observed. Lastly, the measurement is compared with the GSI-Heidelberg~\cite{Andronic:2021erx} and ThermalFIST~\cite{Vovchenko:2019pjl} statistical hadronisation models for Pb--Pb collisions. In pp collisions, a statistical model including the canonical suppression required to describe local quantum-number conservation in small systems (CE-SHMc)~\cite{He:2019tik} is also shown. The statistical models predict an increasing trend with charged-particle multiplicity; however, they overestimate the measured ratio.

Feed-down from excited strange-charm hadrons decaying to non-strange ones could change the trend with multiplicity of the $\mathrm{D_s^+/D^+}$ ratio predicted by different models. Therefore, precise measurements of the production of excited charm states are crucial to interpret the measured $\mathrm{D_s^+/D^+}$ ratio. The first measurement of the $p_\mathrm{T}$-differential $\mathrm{D_{s1}(2536)^+/D_s^+}$ production yield ratio in pp collisions at $\sqrt{s} = 13.6$ TeV is shown in the right panel of Fig.~\ref{fig:mesons}. The ratio is compatible with the $p_\mathrm{T}$-integrated ratio measured by the ALICE Collaboration in Run 2~\cite{ALICE:2024hkk}, and shows a mild increasing trend with $p_\mathrm{T}$. Predictions from \textsc{Pythia 8} and from the TAMU statistical hadronisation model~\cite{He:2019tik} are also shown. The first model, in which the production of the excited strange-charm states was tuned to reproduce Run~2 results, shows good agreement with the data, while the latter underestimates the measured ratio. The GSI-Heidelberg statistical model~\cite{Andronic:2021erx} is in agreement with the $p_\mathrm{T}$-integrated ratio measured in Run~2.

\section{Conclusions}

Charm-hadron production measurements provide key information on charm-quark hadronisation in different collision systems. The $\Lambda_{\mathrm{c}}^{+}/\mathrm{D^{0}}$ ratio measured by ALICE in pp collisions is significantly enhanced with respect to $\mathrm{e^+e^-}$ collisions, indicating that hadronisation is not universal across collision systems. The first measurement of the nuclear modification factor of $\Xi_{\mathrm{c}}^{0}$ baryons in Pb--Pb collisions shows a strong enhancement in the production of strange charm baryons, highlighting the importance of coalescence in charm hadronisation in this system. Measurements of the strange-to-non-strange $\mathrm{D_s^+/D^+}$ charm-meson ratio show that the ratio increases with charged-particle multiplicity in Pb--Pb collisions as compared to pp collisions, and that fragmentation alone, as implemented in \textsc{Pythia 8}, does not fully describe the observed trend. Lastly, the reported measurement of the production of excited charm states provides the first important insights into the role of the feed-down contribution to the $\mathrm{D_s^+/D^+}$ ratio.
\newpage

\bibliographystyle{elsarticle-num}
\bibliography{bibliography}

@article{ALICE:2022exq,
    author = "Acharya, Shreyasi and others",
    collaboration = "ALICE",
    title = "{First measurement of $\mathrm{\Lambda_c^+}$ production down to $p_\mathrm{T}=0$ in pp and p--Pb collisions at $\sqrt{s_\mathrm{NN}}=5.02 TeV$}",
    eprint = "2211.14032",
    archivePrefix = "arXiv",
    primaryClass = "nucl-ex",
    reportNumber = "CERN-EP-2022-261",
    doi = "10.1103/PhysRevC.107.064901",
    journal = "Phys. Rev. C",
    volume = "107",
    number = "6",
    pages = "064901",
    year = "2023"
}

@article{ALICE:2016fzo,
    author = "Adam, Jaroslav and others",
    collaboration = "ALICE",
    title = "{Enhanced production of multi-strange hadrons in high-multiplicity proton-proton collisions}",
    eprint = "1606.07424",
    archivePrefix = "arXiv",
    primaryClass = "nucl-ex",
    reportNumber = "CERN-EP-2016-153",
    doi = "10.1038/nphys4111",
    journal = "Nature Phys.",
    volume = "13",
    pages = "535--539",
    year = "2017"
}

@article{Gladilin:2014tba,
    author = "Gladilin, Leonid",
    title = "{Fragmentation fractions of $c$ and $b$ quarks into charmed hadrons at LEP}",
    eprint = "1404.3888",
    archivePrefix = "arXiv",
    primaryClass = "hep-ex",
    doi = "10.1140/epjc/s10052-014-3250-3",
    journal = "Eur. Phys. J. C",
    volume = "75",
    number = "1",
    pages = "19",
    year = "2015"
}

@article{Skands:2014pea,
    author = "Skands, Peter and Carrazza, Stefano and Rojo, Juan",
    title = "{Tuning PYTHIA 8.1: the Monash 2013 Tune}",
    eprint = "1404.5630",
    archivePrefix = "arXiv",
    primaryClass = "hep-ph",
    reportNumber = "CERN-PH-TH-2014-069, MCNET-14-08, OUTP-14-05P",
    doi = "10.1140/epjc/s10052-014-3024-y",
    journal = "Eur. Phys. J. C",
    volume = "74",
    number = "8",
    pages = "3024",
    year = "2014"
}

@article{Christiansen:2015yqa,
    author = "Christiansen, Jesper R. and Skands, Peter Z.",
    title = "{String Formation Beyond Leading Colour}",
    eprint = "1505.01681",
    archivePrefix = "arXiv",
    primaryClass = "hep-ph",
    reportNumber = "COEPP-MN-15-1, LU-TP-15-16, MCNET-15-09, COEPP-MN-15-1, LU-TP-15-16, MCNET-15-09",
    doi = "10.1007/JHEP08(2015)003",
    journal = "JHEP",
    volume = "08",
    pages = "003",
    year = "2015"
}

@article{Minissale:2020bif,
    author = "Minissale, Vincenzo and Plumari, Salvatore and Greco, Vincenzo",
    title = "{Charm hadrons in pp collisions at LHC energy within a coalescence plus fragmentation approach}",
    eprint = "2012.12001",
    archivePrefix = "arXiv",
    primaryClass = "hep-ph",
    doi = "10.1016/j.physletb.2021.136622",
    journal = "Phys. Lett. B",
    volume = "821",
    pages = "136622",
    year = "2021"
}

@article{Song:2018tpv,
    author = "Song, Jun and Li, Hai-hong and Shao, Feng-lan",
    title = "{New feature of low $p_{T}$ charm quark hadronization in $pp$ collisions at $\sqrt{s}=7$ TeV}",
    eprint = "1801.09402",
    archivePrefix = "arXiv",
    primaryClass = "hep-ph",
    doi = "10.1140/epjc/s10052-018-5817-x",
    journal = "Eur. Phys. J. C",
    volume = "78",
    number = "4",
    pages = "344",
    year = "2018"
}

@article{Beraudo:2023nlq,
    author = "Beraudo, Andrea and De Pace, Arturo and Pablos, Daniel and Prino, Francesco and Monteno, Marco and Nardi, Marzia",
    title = "{Heavy-flavor transport and hadronization in pp collisions}",
    eprint = "2306.02152",
    archivePrefix = "arXiv",
    primaryClass = "hep-ph",
    doi = "10.1103/PhysRevD.109.L011501",
    journal = "Phys. Rev. D",
    volume = "109",
    number = "1",
    pages = "L011501",
    year = "2024"
}

@article{Ebert:2011kk,
    author = "Ebert, D. and Faustov, R. N. and Galkin, V. O.",
    title = "{Spectroscopy and Regge trajectories of heavy baryons in the relativistic quark-diquark picture}",
    eprint = "1105.0583",
    archivePrefix = "arXiv",
    primaryClass = "hep-ph",
    reportNumber = "HU-EP-11-21",
    doi = "10.1103/PhysRevD.84.014025",
    journal = "Phys. Rev. D",
    volume = "84",
    pages = "014025",
    year = "2011"
}

@article{He:2019tik,
    author = "He, Min and Rapp, Ralf",
    title = "{Charm-Baryon Production in Proton-Proton Collisions}",
    eprint = "1902.08889",
    archivePrefix = "arXiv",
    primaryClass = "nucl-th",
    doi = "10.1016/j.physletb.2019.06.004",
    journal = "Phys. Lett. B",
    volume = "795",
    pages = "117--121",
    year = "2019"
}

@article{Zhao:2024ecc,
    author = "Zhao, Jiaxing and Aichelin, Joerg and Gossiaux, Pol Bernard and Ozvenchuk, Vitalii and Werner, Klaus",
    title = "{Heavy-flavor hadron production in relativistic heavy ion collisions at energies available at BNL RHIC and at the CERN LHC in the EPOS4HQ framework}",
    eprint = "2401.17096",
    archivePrefix = "arXiv",
    primaryClass = "hep-ph",
    doi = "10.1103/PhysRevC.110.024909",
    journal = "Phys. Rev. C",
    volume = "110",
    number = "2",
    pages = "024909",
    year = "2024"
}

@article{Vovchenko:2019pjl,
    author = "Vovchenko, Volodymyr and Stoecker, Horst",
    title = "{Thermal-FIST: A package for heavy-ion collisions and hadronic equation of state}",
    eprint = "1901.05249",
    archivePrefix = "arXiv",
    primaryClass = "nucl-th",
    doi = "10.1016/j.cpc.2019.06.024",
    journal = "Comput. Phys. Commun.",
    volume = "244",
    pages = "295--310",
    year = "2019"
}

@article{Andronic:2021erx,
    author = {Andronic, Anton and Braun-Munzinger, Peter and K{\"o}hler, Markus K. and Mazeliauskas, Aleksas and Redlich, Krzysztof and Stachel, Johanna and Vislavicius, Vytautas},
    title = "{The multiple-charm hierarchy in the statistical hadronization model}",
    eprint = "2104.12754",
    archivePrefix = "arXiv",
    primaryClass = "hep-ph",
    doi = "10.1007/JHEP07(2021)035",
    journal = "JHEP",
    volume = "07",
    pages = "035",
    year = "2021"
}

@article{He:2019vgs,
    author = "He, Min and Rapp, Ralf",
    title = "{Hadronization and Charm-Hadron Ratios in Heavy-Ion Collisions}",
    eprint = "1905.09216",
    archivePrefix = "arXiv",
    primaryClass = "nucl-th",
    doi = "10.1103/PhysRevLett.124.042301",
    journal = "Phys. Rev. Lett.",
    volume = "124",
    number = "4",
    pages = "042301",
    year = "2020"
}

@article{Bierlich:2018xfw,
    author = {Bierlich, Christian and Gustafson, G{\"o}sta and L{\"o}nnblad, Leif and Shah, Harsh},
    title = "{The Angantyr model for Heavy-Ion Collisions in PYTHIA8}",
    eprint = "1806.10820",
    archivePrefix = "arXiv",
    primaryClass = "hep-ph",
    reportNumber = "LU-TP-18-19, LU-TP 18-19, MCnet-18-12",
    doi = "10.1007/JHEP10(2018)134",
    journal = "JHEP",
    volume = "10",
    pages = "134",
    year = "2018"
}

@article{ALICE:2024hkk,
    author = "Acharya, Shreyasi and others",
    collaboration = "ALICE",
    title = "{First measurement of $\mathrm{D_{s1}(1^+)(2536)^+}$ and $\mathrm{D_{s2}^*(2^+)(2573)^+}$ production in proton-proton collisions at $\sqrt{s} = 13{\,}$TeV at the LHC}",
    eprint = "2409.11938",
    archivePrefix = "arXiv",
    primaryClass = "hep-ex",
    reportNumber = "CERN-EP-2024-233",
    doi = "10.1103/PhysRevD.111.112005",
    journal = "Phys. Rev. D",
    volume = "111",
    number = "11",
    pages = "112005",
    year = "2025"
}



\end{document}